# Towards the Standardization of Non-orthogonal Multiple Access for Next Generation Wireless Networks

Yan Chen, Alireza Bayesteh, Yiqun Wu, Bin Ren, Shaoli Kang, Shaohui Sun,
Qi Xiong, Chen Qian, Bin Yu, Zhiguo Ding, Sen Wang, Shuangfeng Han,
Xiaolin Hou, Hao Lin, Raphael Visoz, and Razieh Razavi[1]

Abstract

Non-orthogonal multiple access (NoMA) as an efficient way of radio resource sharing can root back to the network information theory. For generations of wireless communication systems design, orthogonal multiple access (OMA) schemes in time, frequency, or code domain have been the main choices due to the limited processing capability in the transceiver hardware, as well as the modest traffic demands in both latency and connectivity. However, for the next generation radio systems, given its vision to connect everything and the much evolved hardware capability, NoMA has been identified as a promising technology to help achieve all the targets in system capacity, user connectivity, and service latency. This article will provide a systematic overview of the state-of-the-art design of the NoMA transmission based on a unified transceiver design framework, the related standardization progress, and some promising use cases in future cellular networks, based on which the interested researchers can get a quick start in this area.

**Key words**: Non-orthogonal multiple access (NoMA), unified transceiver design framework, NoMA enabled grant-free transmission, NoMA enabled collaborative communications

# 1 Introduction of Multiple Access

Radio resource is the medium in wireless communications to transmit data information from one device to another. The fundamental physical radio resource is time and frequency, which is usually interpreted as physical degrees of freedom to transmit data. The problem of multiple access comes when multiple users are going to be served with limited (or scarce) degrees of freedom in the radio resource.

Yan Chen, Alireza Bayesteh, and Yiqun Wu are with Huawei Technologies Co., Ltd.; Bin Ren, Shaoli Kang and Shaohui Sun are with China Academy of Telecommunications Technology (CATT); Qi Xiong, Chen Qian, and Bin Yu are with Samsung Research Institute China (SRC-B); Zhiguo Ding is with Lancaster University; Sen Wang and Shuangfeng Han are with China Mobile Research Institute; Xiaolin Hou is with DOCOMO Beijing Labs; Hao Lin and Raphael Visoz are with Orange Labs; Razieh Razavi is with Vodafone.

## 1.1 Orthogonal multiple access

It is intuitive to consider dividing the available degrees of freedom in an orthogonal way so that each user's transmission will not interfere with another. This way of orthogonal multiple access (OMA) design starts from the very early generation of digital cellular communications such as the second generation (2G) Global System for Mobile Communications (GSM), till the recent fourth generation (4G) Long Term Evolution (LTE). However, each generation has different ways to divide the degrees of freedom, as illustrated in Figure 1. In time division multiple access (TDMA), time is partitioned into time slots each serving an digital data stream in a round-robin fashion; while in frequency division multiple access (FDMA), the available spectrum is partitioned into non-overlapped frequency sub-bands each accommodating one digital data stream. Orthogonal frequency division multiple access (OFDMA) is a multi-carrier multiple access scheme based on the orthogonal frequency division multiplexing (OFDM) waveform, which enables tight and orthogonal frequency-domain packing of the subcarriers with a subcarrier spacing inverse to the symbol duration. In light of this, the time and frequency plane with OFDMA are divided into two-dimensional raster, each transmitting a modulated symbol that belongs to one data stream.

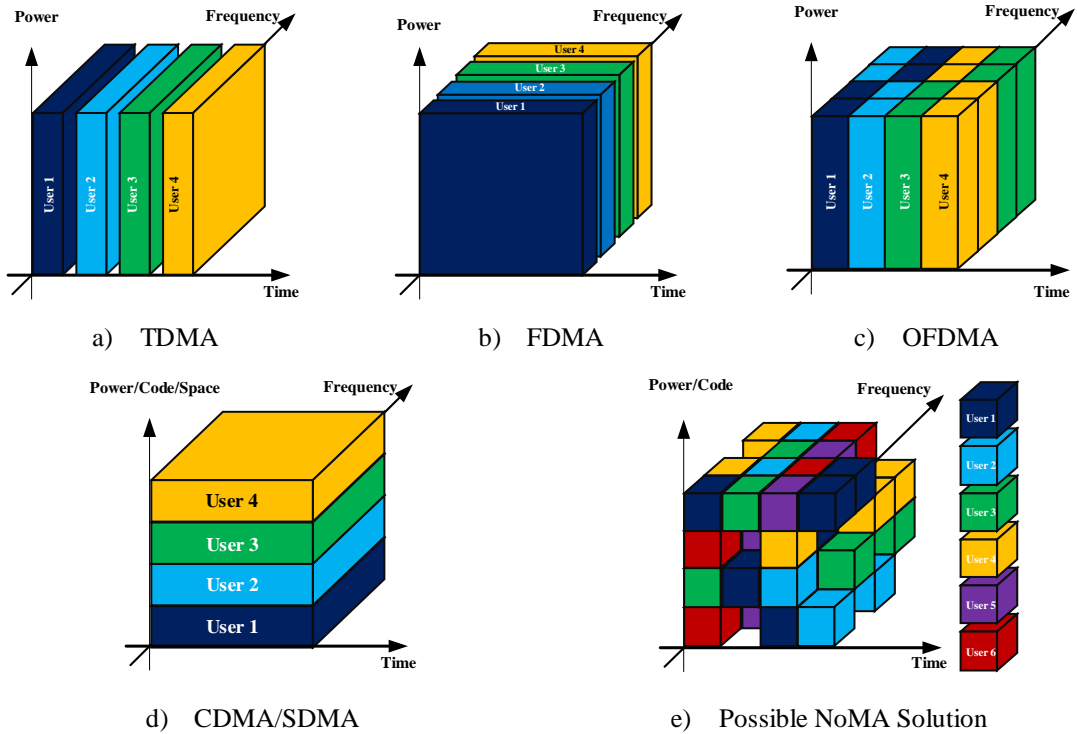

a)  TDMA            b)  FDMA            c)  OFDMA

d)  CDMA/SDMA            e)  Possible NoMA Solution

**Figure 1 Illustrative example of different multiple access schemes.**

On top of time and frequency resources, more degrees of freedom can be created by introducing the code domain or spatial domain resource together with the corresponding signaling processing technologies. Code division multiple access (CDMA) is an example in which some user specific code signatures are used to spread the modulated symbol by a factor of length N, which is also known as the processing gain. Note that the code signatures can be orthogonal to each other, in which case, CDMA can also be taken as one type of OMA schemes and the number of users that can be simultaneously supported is less than or equal to N. However, it is also possible to tradeoff orthogonality for higher system throughput in order to accommodate more users simultaneously. In this sense, CDMA can also be considered as a type of non-orthogonal multiple access schemes. Similarly, spatial division multiple access (SDMA) can either be orthogonal or non-orthogonal, depending on which precoding method is applied.

The benefit of the OMA schemes is clear, i.e., simplifying the transceiver design and avoiding any intra-cell co-channel interference. However, the limitations are obvious too. First, the number of users that can be served simultaneously is limited strictly by the pool of the radio resource. Second, careful user scheduling with dedicated feedback channels at the expense of signaling overhead is needed to guarantee the orthogonality.

### 1.2  Non-orthogonal multiple access

Compared to OMA, non-orthogonal multiple access (NoMA) opens the horizon for a new angle of thinking. In particular, by relaxing the constraint of orthogonal radio resource allocation, the user scheduling problem constrained by the limited time and bandwidth resources is no longer a binary selection, but the optimization of joint power, code signature, and receiver design. As it has long been predicted by the network information theory [1], the total number of users served as well as the overall capacity of the system can be greatly improved in a NoMA network as compared with that of OMA network, especially when advanced multi-user detection algorithms are applied. Moreover, due to the non-orthogonal nature, the requirement of precise channel feedback and scheduling for multi-user multiplexing is thus reduced, or even removed in some scenarios.

A generic example of non-orthogonal multiple access is described in figure 1-e), in which different users are multiplexed in three domains of time, frequency and power/code, which means the users are not orthogonal on any of the domains alone. However, by applying appropriate code design and time/frequency occupation patterns, users can be efficiently decoded/separated while a better overall performance can be achieved compared to OMA.

The rest of the paper will elaborate the recent progress of NoMA standardization in 3GPP, especially in the UL, and the basic features of NoMA transceivers based on a unified framework. The primary goal is to provide a systematic way for the interested researchers to get a quick understanding of the state-of-the-art design principles for NoMA transceivers. Two interesting application examples of NoMA enabled UL grant-free transmission for small packets [2] and NoMA enabled open-loop collaborative transmission in DL [3] are then given to further elaborate the benefit of NoMA. Conclusions and challenges are also outlined at the end of the paper to shed light on possible future works.

## 2  NoMA Standardization Progress in 3GPP

The design of 5G radio networks is targeting towards higher capacity, larger connectivity, and lower latency, which shall not only provide better user experience for enhanced mobile broad band (eMBB) services, but also connect to new vertical industries and new devices, creating advanced application scenarios such as massive Machine Type Communication (mMTC) and Ultra Reliable Low Latency Communication (URLLC) services. The mMTC application scenario targets to support a massive number of devices simultaneously, while the URLLC scenario enables mission critical transmissions with ultra high reliability and ultra low latency. Towards these goals and among all components in the radio link design, NoMA has attracted great attention across both academia and industry [2-15].

For instance, the application of NoMA in eMBB is expected to increase the multi-user capacity, provide better fairness against the near-far effect and improve user experience in ultra dense networks. While for the URLLC scenario, the application of NoMA can enable ultra reliable link quality when contention based grant-free transmission is applied to achieve ultra low latency. It is also important to point out that the application of NoMA enables efficient multiplexing of URLLC and eMBB services to further improve resource utilization. Finally for the mMTC scenario, NoMA is by far the most competitive solution to address the massive connectivity issue together with the large coverage requirement. In the following, we will elaborate the recent NoMA standardization progress in 3GPP for both DL and UL, respectively.

## 2.1 DL NoMA Standardization

The recent study of NoMA in 3GPP starts in LTE Release-13 under the name Multi-User Superposed Transmission (MUST), mainly focusing on DL transmission. The MUST schemes can be categorized into three categories [5]. In MUST Category 1, coded bits of two or more co-scheduled users are independently mapped to component constellation symbols but the composite constellation does not have Gray mapping. In MUST Category 2, coded bits of two or more co-scheduled users are jointly mapped to component constellations and then the composite constellation has Gray mapping. In MUST Category 3, coded bits of two or more co-scheduled users are directly mapped onto the symbols of a composite constellation.

It is expected that in the future, MUST schemes, possibly with some new features will be considered in 5G. The evolved techniques may also be combined with the beam management techniques designed in the scenario with a large number of transmit and/or receive antennas.

## 2.2 UL NoMA Standardization

In 3GPP Release-14 study for New Radio (NR) system design, 15 NoMA schemes have been proposed, mainly targeting UL transmissions to support massive connectivity and to enable the newly defined grant-free transmission procedures with low latency and high reliability. A full list of schemes and the corresponding 3GPP contributions describing the schemes are given below.

- Sparse code multiple access (SCMA)
- Multi-user shared access (MUSA)
- Low code rate spreading
- Frequency domain spreading
- Non-orthogonal coded multiple access (NCMA)
- Non-orthogonal multiple access (NOMA)
- Pattern division multiple access (PDMA)
- Resource spread multiple access (RSMA)
- Interleave-Grid Multiple Access (IGMA)
- Low density spreading with signature vector extension (LDS-SVE)
- Low code rate and signature based shared access (LSSA)
- Non-orthogonal coded access (NOCA)
- Interleave Division Multiple Access (IDMA)
- Repetition division multiple access (RDMA)
- Group Orthogonal Coded Access (GOCA)

It was hard to reach final decision on the down selection of the schemes in the limited study period, however, comprehensive link-level and system-level simulations have been performed by different companies to justify the gain of NoMA over OFDMA which is used as an OMA baseline. From the comprehensive simulation campaign, it has been agreed that for the evaluated scenarios, significant benefit of NoMA can be observed in terms of uplink link-level sum throughput and overloading capability, as well as system capacity enhancement in terms of supported packet arrival rate (PAR) at a given system outage level such as 1% packet drop rate (PDR).

Moreover, a new study item (SI) has been approved to continue studying uplink NoMA schemes in Release-15. The content of the SI will cover transmitter side signal processing, multi-user receiver design and complexity analysis, NoMA related procedures such as HARQ, link adaptation, power/signature allocation, etc.. In addition, this new study will also include more evaluation work continued from performance metrics identified in Release-14 and for all scenarios including eMBB, URLLC, and mMTC, taking into consideration more realistic modeling of non-ideal impairment at both the transmitter and receivers side, such as potential PAPR issue, channel estimation error, power control accuracy, and NoMA signature collision.

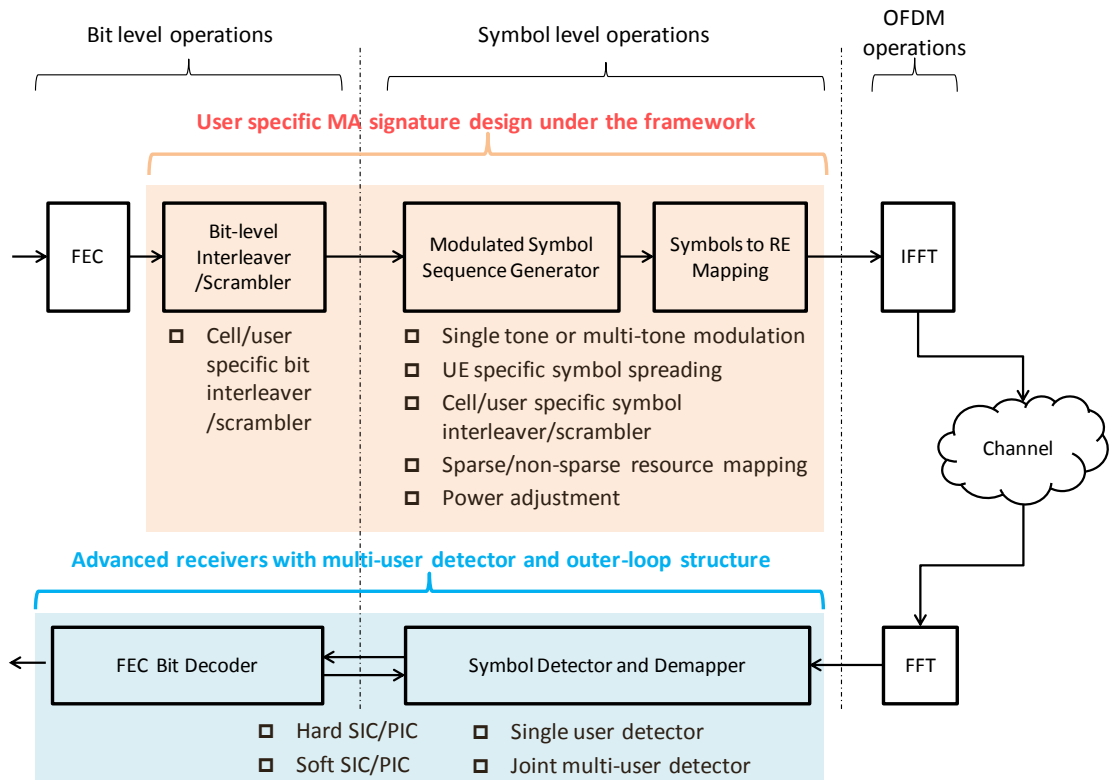

**Figure 2 Unified framework for UL NoMA design.**

# 3  Basic Features and Unified Design Framework

In this section, we shall introduce the basic features and design principles of NoMA schemes based on a unified transceiver framework. The discussion will mainly focus on UL NoMA where the random channel is applied to each user before the multiple data signals from different users are multiplexed together. Such a property prevents the design of joint constellation with superposition in advance as in MUST category 2 and 3 and calls for design from per user (or per layer) aspect that can accommodate the randomness brought by the user specific channels.

In general, each UL NoMA scheme at the transmitter side by nature tries to map the information bits to the available transmission resources by some user-specific operations to facilitate decoding of the superposed multi-user data at the receiver side with reasonable complexity. These operations can involve both the bit domain and symbol domain signal processing, which can be unified in a general framework as shown in Figure 2. The differences between the NoMA schemes will then be reflected in *NoMA signature* design at the transmitter side by configuring all or a subset of these component blocks.

### 3.1　Transmitter Side Building Blocks

Following the unified framework in Figure 2, each NoMA signature is a combination of different components along the framework at the transmitter side. Since the forward error correction coding (FEC) and OFDM operation blocks are common for all the NoMA schemes, the unique features of any proposed NoMA transmitter design are thus characterized by the three component blocks: 1) bit-level interleaver and/or scramble; 2) modulated symbol sequence generator; and 3) symbol to resource element (RE) mapping. Within each of these three component blocks, there are further options to be configured, as illustrated below.

- *Bit-level interleaver and/or scrambler*

  In the current LTE system, both user-specific and cell-specific bit scrambling can be applied. The main benefit of having interleaving or scrambling is to randomize the inter-user/inter-cell interference. Then it is interesting to find out whether user-specific bit interleaver could bring extra benefits on top of the user-specific bit scrambling, and whether it could further facilitate symbol domain NoMA signature design.

- *Modulated symbol sequence generator*

  This block converts the sequence of input coded bits to a sequence of symbols to be mapped to the REs that transmitted over the air. The details of how the streams of bits are converted to the streams of symbols can be configured to be user-specific. This block includes different ways of modulation, spreading, and interleaver/scrambler that can be configured by each user to construct its own NoMA signatures. For NoMA signatures that include the feature of symbol-level spreading, the spreading length, spreading type (modulation dependent or not), and spreading signatures/codebooks can be designed to facilitate the multi-user detection at the receiver side. Besides the configured symbol-level spreading, symbol-level interleaving/scrambling may be configured by each user as another dimension to help distinguish user and/or randomize interference. Moreover, power adjustment as a power domain feature can be configured with and without the other spreading/scrambler features.

- *Symbol to RE mapping*

  Symbol-to-RE mapping can be non-sparse (i.e. all symbols take all available resource elements), or sparse (i.e. symbols occupy only a portion of the available resources elements). In the latter case, the sparsity level and the symbol-to-RE mapping pattern can be configured to be user-specific to facilitate the multi-user detection. Note that sparse symbol-to-RE mapping can also be interpreted as part of spreading in the sense that the actual REs for the group of information bits are expanded by adding zero tones.

One of the key tasks of the NoMA SI in Release-15 is to figure out how to configure each of these building blocks so that different performance metrics such as block error rate, connection density, throughput, PAPR, energy efficiency, can be achieved for each of the eMBB, URLLC, and mMTC scenarios with scenario specific requirements and assumptions.

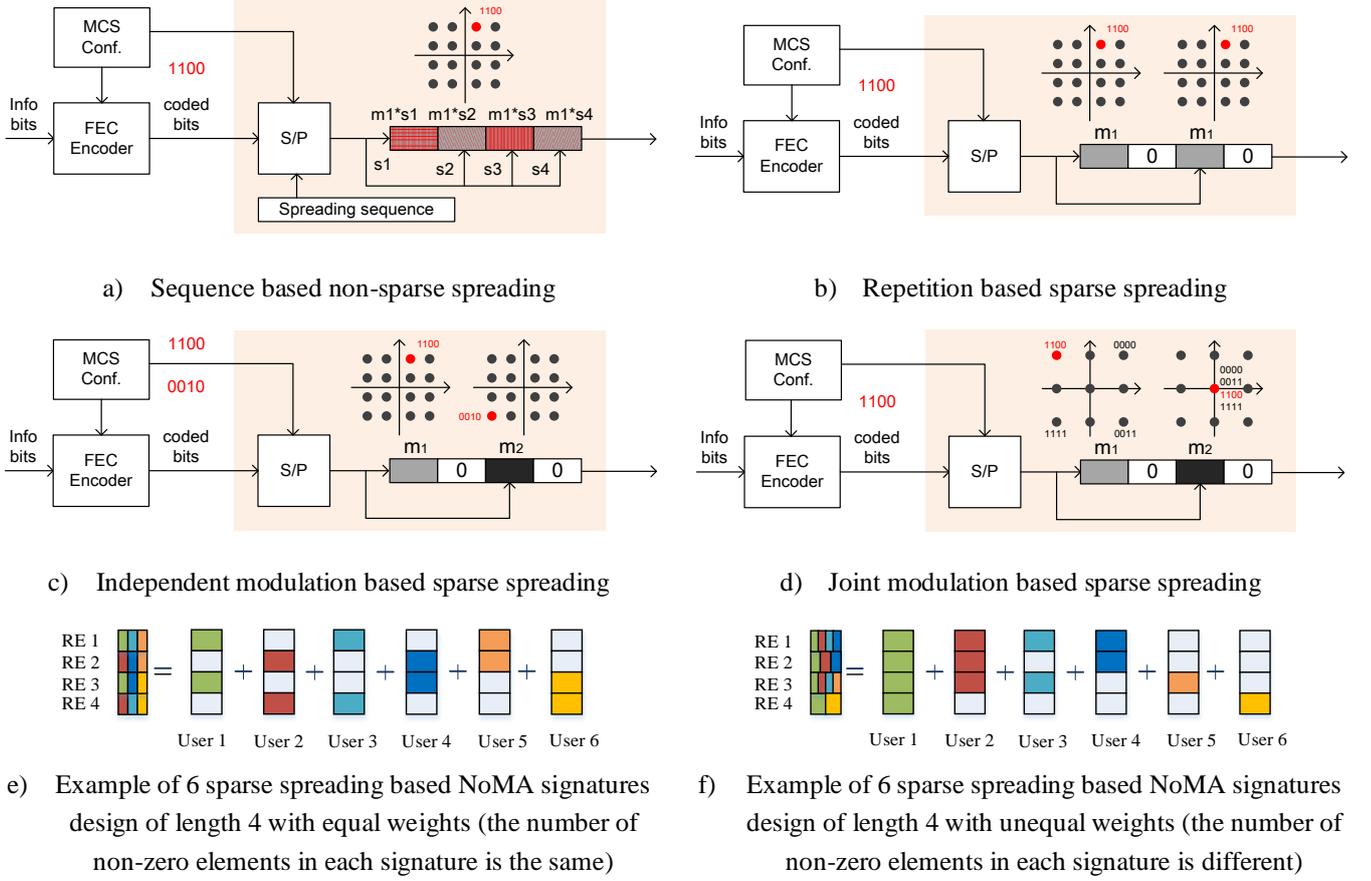

a) Sequence based non-sparse spreading
b) Repetition based sparse spreading
c) Independent modulation based sparse spreading
d) Joint modulation based sparse spreading
e) Example of 6 sparse spreading based NoMA signatures design of length 4 with equal weights (the number of non-zero elements in each signature is the same)
f) Example of 6 sparse spreading based NoMA signatures design of length 4 with unequal weights (the number of non-zero elements in each signature is different)

**Figure 3 Illustration of different spreading features.**

- *Example of configurations*

   Having discussed different options in each component block, Figure 3 shows some examples of configured features at the symbol level with different types of spreading and RE mapping.

1) *Example configuration 1*: *Sequence based non-sparse spreading*. In this configuration, per symbol modulation is applied together with sequence based spreading and non-sparse symbol-to-RE mapping, as shown in Figure 3-a). The optimization variables in this configuration mainly lie in the design of the low correlation spreading sequence [6].

2) *Example configuration 2*: *Repetition based sparse spreading*. In this configuration, per symbol modulation is applied together with repetition based spreading and sparse symbol-to-RE mapping, as shown in Figure 3-b). The optimization variables in this configuration lie in the choices of spreading length and sparsity patterns with equal or unequal weights [7], as shown in Figure 3-e) and 3-f), respectively.

3) *Example configuration 3*: *Independent modulation based sparse spreading*. In this configuration, per symbol modulation with independent bit groups is applied and interleaved zero tones to have sparse symbol-to-RE mapping, as shown in Figure 3-c). The optimization variables in this configuration mainly depend on the symbol interleaver design to introduce zeros into a block of non-zero symbols with user-specific sparsity patterns. By selecting different levels of sparsity, this configuration can have the flexibility to trade between larger channel coding gain and less inter-user interference [8].

4) *Example configuration 4: Joint modulation based sparse spreading.* In this configuration, joint multi-symbol modulation with good distance properties (Euclidean and/or Product) among the points in the overall multi-symbol constellation is applied together with sparse symbol-to-RE mapping, as shown in Figure 3-d). The optimization variables in this configuration mainly lie in the joint multi-symbol constellation design to maximize the coding/shaping gain compared with per symbol modulation and spreading, and also in the selection of spreading length and sparsity patterns to adaptively trade between higher signal diversity and lower inter-user interference based on scenario requirement [9]. Note that in this configuration, similar to example configurations 1 and 2, code domain interference suppression techniques are applied, and similar to example configurations 2 and 3, sparse symbol-to-RE mapping is introduced for supporting more superposed users with affordable receiver complexity. In addition, this configuration exploits the modulation domain optimization, which can further improve spectrum efficiency.

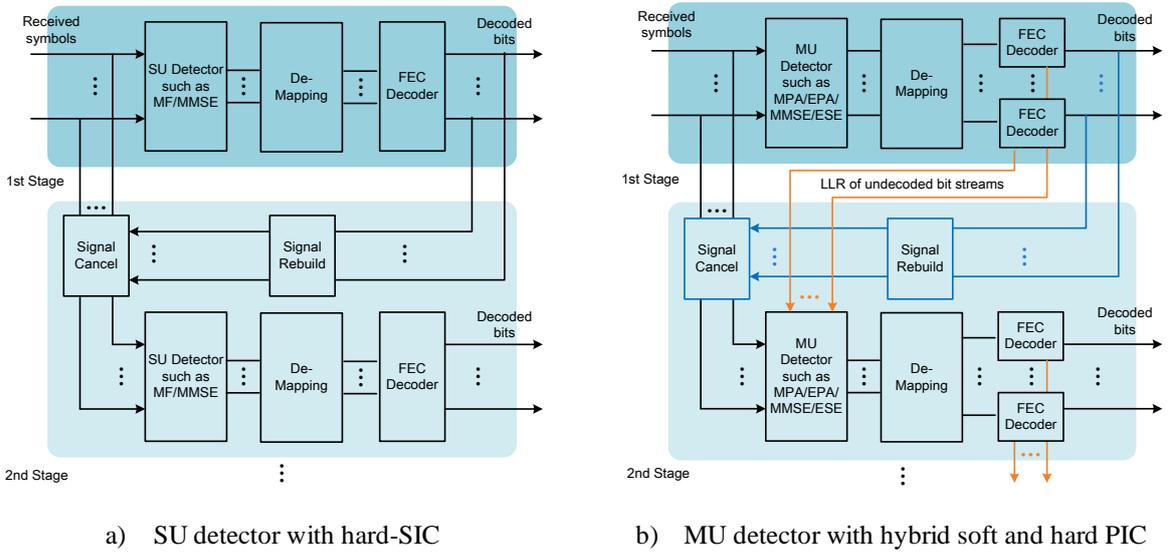

a) SU detector with hard-SIC  b) MU detector with hybrid soft and hard PIC

**Figure 4 Illustration of typical receiver structures for NoMA multi-user detection.**

## 3.2  Receiver Side Structures

In theory, the optimal multi-user receiver needs a fully joint design of symbol-level detection and bit-level FEC decoding, which however, has prohibitively high complexity for practical implementation. The other extreme is to completely separate the two operations, which is simple but may suffer from severe performance degradation as compared with the joint design. In practical systems, one can come up with a more realistic design where a unified Turbo-like outer-loop structure is adopted to allow iterations between the symbol detector and the FEC decoder. This Turbo-like outer loop structure is briefly illustrated in Figure 2.

Both the single-user (SU) detection and joint multi-user (MU) detection can be applied to the *Symbol Detector and Demapper* block. Here SU detection means that a single user's signal is detected treating other users' signals as noise, while MU detection means multiple users' signals are decoded together and decoding one user's data uses information from the signal of other users. Classic SU detector includes algorithms such as the matched filter (MF) and SU minimum mean square error (MMSE) estimator, while a typical MU detector includes the maximum a posterior probability (MAP) algorithm, maximum likelihood (ML) algorithm, message passing algorithm (MPA) [10], expectation propagation algorithm (EPA) [11], as well as MU MMSE estimator [6] and elementary signal estimator (ESE) [12], etc. Note that in the case of spreading, the MAP/ML and MMSE can be done either in a block based way (perform the detection method jointly for the symbols within one spread block) or in a chip based manner, e.g., chip-by-chip MAP [8], in which the latter has lower complexity.

In particular, MAP and ML refer to the optimal receivers based on the maximum a posterior probability decision criterion and maximum likelihood decision criterion, respectively. With a uniform prior probability, MAP is equivalent to ML. MPA can be considered as a kind of approximation of MAP/ML detector by introducing the message passing procedure on the factor graph to replace the direct probability calculation [9], where the sparsity in NoMA signature can further reduce its complexity compared with ML detection but keep similar performance. EPA takes a next step to reduce complexity by iteratively approximating the posterior probability distribution as a Gaussian distribution, thus changing the message passing procedure to update means and variances only, whose complexity grows linearly with the number of users.

On top of all these detectors, successive interference cancelation (SIC) can be applied in the outer-loop structure with either hard-SIC or soft-SIC operations. Specifically, for hard-SIC operation, only the successfully decoded signals are cancelled and no soft information is fed from the FEC decoder back to the symbol detector for the unsuccessfully decoded data streams, as shown in Figure 4-a). For soft-SIC, on the contrary, soft information from the FEC decoder such as extrinsic log-likelihood ratio (LLR) is fed back to the symbol detector as the prior information for the next round of detection. Note that for the joint MU detector, parallel interference cancellation (PIC) instead of SIC can be applied to reduce decoding latency. Hard-PIC and soft-PIC can be combined in the sense that for users with decoded bits, reconstruction and cancellation are performed, while for those users with non-decoded bits, soft LLR can be fed back as inputs for the symbol detector, as shown in Figure 4-b).

# 4  Use Cases in Cellular Networks

## 4.1  NoMA enabled grant-free transmission

Grant-free transmission is a mechanism that eliminates the dynamic scheduling request (SR) and grant signaling overhead for uplink data transmission and a user can transmit uplink data in an "arrive-and-go" manner [2]. The benefits of grant-free transmission include overhead reduction, latency reduction, and energy saving especially at the user side with longer sleeping time.

With grant-free transmission, contention is usually allowed to increase the system resource utilization, i.e., the users may transmit on the same time and frequency resource as there is no coordination from the base station. In this case, NoMA based grant-free transmission will show its advantage as a solution for contention resolution with high reliability, since they are designed with high overloading capability. The design of NoMA based grant-free transmission has been proposed and discussed during Release-14 NR Study, in which NoMA signatures are taken as part of grant-free resource besides the traditional physical resource such as time and frequency resource. Prior to transmission, a user can either randomly select one NoMA signature to transmit from a given resource pool, or transmit with a pre-configured NoMA signature. Then in each of the contention region (the basic unit of physical resource for grant-free transmission), multiple NoMA signatures from different users will be multiplexed, as shown in Figure 5-b). User specific pilots are assumed for user activity identification and channel estimation.

One design challenge for NoMA based grant-free transmission is to deal with the potential signature collision, which will happen in the case of random signature selection, or when the number of potential users is much larger than the pool size of the NoMA signatures. This demands the consideration of collision robustness in the component configuration at the transmitter side and the selection of collision-resilient MU detectors at the receiver side. Moreover, with more users multiplexed together, how to guarantee good user detection performance and channel estimation quality offered by the extended pool of pilots is another interesting topic to explore [13].

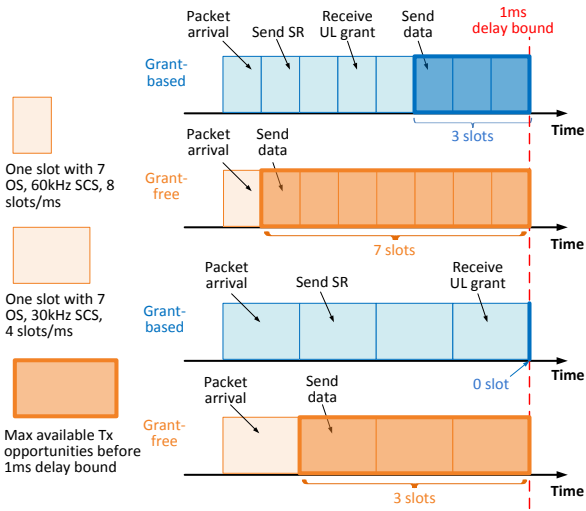

a) Illustrative example of how grant-free URLLC can have more data repetition/retransmission opportunities

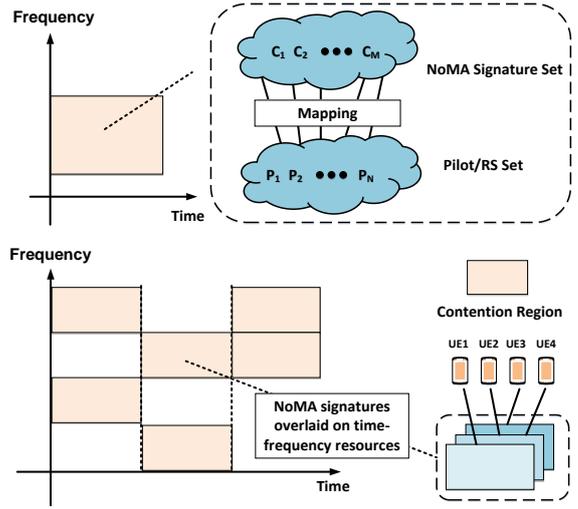

b) Illustrative example of how NoMA enabled grant-free network works.

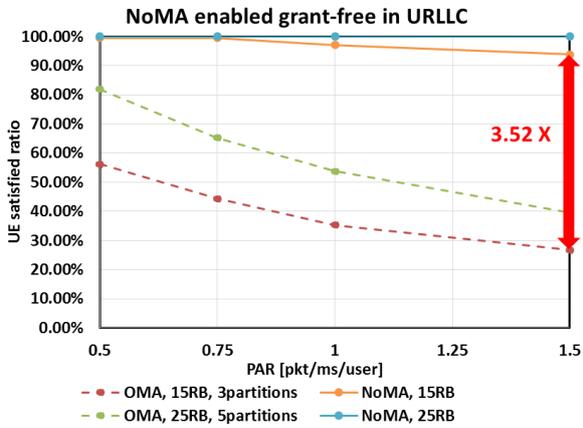

| PAR @1500 | OMA | NoMA | Gain (times) |
|---|---|---|---|
| 15 RB (3 partitions for OMA) | 0.2667 | 0.9381 | 3.52 |
| 25 RB (5 partitions for OMA) | 0.3875 | 0.9952 | 2.57 |

c) Example performance gain of NoMA enabled grant-free over OFDMA based grant-free in terms of the ratio of satisfied users (successfully delivering more than 99.99% of its total packets each within 1ms) among all users at given PAR in the URLLC scenario.

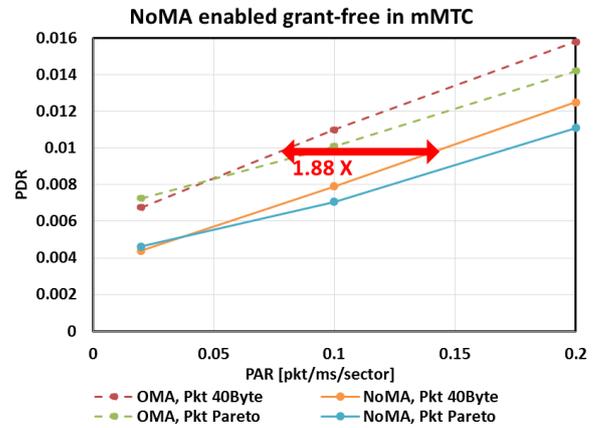

| PAR @PDR 1% | OMA | NoMA | Gain (times) |
|---|---|---|---|
| Pkt size = 40 Bytes | 0.08 | 0.15 | 1.88 |
| Pkt size = Pareto | 0.1 | 0.175 | 1.75 |

d) Example performance gain of NoMA enabled grant-free over OFDMA based grant-free in terms of supported PAR at given PDR (e.g., 1%) in the mMTC scenario with extreme coverage case (maximum coupling loss (MCL) of 164dB) considered.

**Figure 5 Illustration and benefit of NoMA enabled grant-free network.**

Some example system-level simulation results are shown in Figure 5-c) and 5-d) for URLLC and mMTC scenarios, respectively. The attributes of simulation methodology including physical layer abstraction are delineated in [14]. In each figure, NoMA (taking SCMA as an example) enabled grant-free transmission is compared with OFDMA based grant-free transmission with the same parameter settings (e.g., the same traffic model and path-loss model, the same total available bandwidth, and the same average power per user). It can be observed from the figures that with NoMA design, at the same PAR, the ratio of satisfied users (i.e., both the latency and reliability requirement are met) in URLLC can be significantly increased. The smaller the total bandwidth is, the larger the gain is. And in the mMTC case, even with some users in very deep coverage, NoMA enabled grant-free transmission could still bring about 88% gain at 1% system PDR.

### 4.2 NoMA enabled collaborative communications

One of the solutions for interference coordination in wireless networks is cooperation among transmit points (TPs) which is also known as coordinated multi-point (CoMP) transmission. Most proposed CoMP schemes in 3GPP up to Release-14 are closed-loop precoding based on the short-term channel state information (CSI) feedback from users to the cooperating TPs. CSI feedback can be quite challenging in the future networks due to an excessive number of users and TPs especially for Ultra Dense Networks (UDNs) where a user is seen by a large number of TPs.

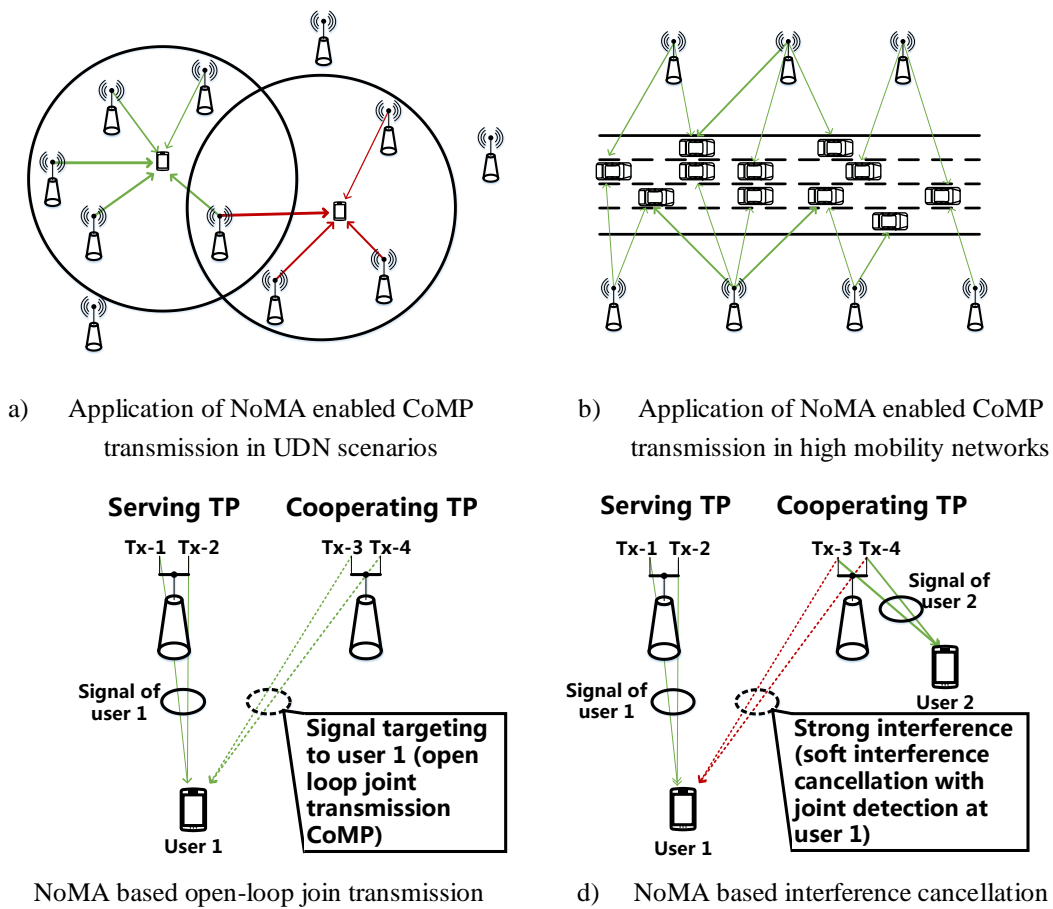

a) Application of NoMA enabled CoMP transmission in UDN scenarios

b) Application of NoMA enabled CoMP transmission in high mobility networks

c) NoMA based open-loop join transmission

d) NoMA based interference cancellation

**Figure 6 Illustration of NoMA enabled open-loop collaborative communications**

NoMA with inter-TP layer assignment through a central scheduler can provide an open-loop CoMP solution without the knowledge of short-term multi-TP CSI [3]. It can bring two main advantages to the system, namely 1) dramatic reduction of the overhead caused by dynamic multi-TP CSI feedback, and 2) significant increase of the robustness to channel aging. More specifically, in an open-loop CoMP solution enabled by NoMA, different NoMA signature sets are assigned to different TP antennas. Each transmit antenna uses a specific NoMA signature set to multiplex UEs. Terminals jointly detect the signals from multiple TPs within their CoMP collaborative cluster. The cluster size depends on the network topology. On the other hand, a TP may serve multiple users if they have overlapped CoMP clusters. It enables user-centric CoMP via NoMA signature allocation across multiple TPs. Multiple links to a user can facilitate soft handover across a UDN network or high mobility networks such as in V2X, as shown in Figure 6-a) and 6-b), where frequent handover becomes a technical challenge.

Note that a neighboring TP can be either a cooperating TP or an interfering TP. In the cooperating TP case of Figure 6-c), the signal from a neighboring TP targets the same user and the open-loop joint-transmission is performed to improve the coverage for cell-edge users. An alternative CoMP solution is to use the NoMA receiver for soft interference cancelation, as shown in Figure 6-d). Moreover, the mode of joint transmission and soft interference cancellation through MU detector can be combined to improve both the cell edge and cell average throughputs especially in a UDN network.

## 5  Summary and Future Directions

In summary, NoMA is an attractive solution to boost the system capacity by accommodating more users at the same time/frequency resource, reduce system latency caused by scheduling and queueing to guarantee inter-user orthogonality, as well as to relax the dependency on precise channel state information and feedback quality. In particular, for UL, NoMA enabled grant-free is a competitive solution for small packet transmission in many scenarios including mMTC, URLLC, and eMBB, while for DL, besides MUST, NoMA enabled open-loop CoMP solution can be attractive in UDN and high mobility networks to help boost cell edge performance and solve the frequent handover issues.

In the coming study of 3GPP NoMA SI, more works will be dedicated to the comprehensive evaluations of the various candidate schemes based on the unified framework to better understand the commonality and differentiation of different schemes, and thus to find the recommended configurations for different target scenarios. Moreover, as other technologies are evolving in parallel in 3GPP, the study of how these radio technologies can be integrated with NoMA shall be carried out. As one example, the integration of NoMA with (massive) MIMO has been raised in literatures [15]. Recent studies have demonstrated the transmission features of massive MIMO, such as geometric channel correlation and the use of finite resolution analog beamforming, facilitate the implementation of NoMA in massive MIMO scenarios, and improve the spectral efficiency significantly compared to the OMA scenarios.

**YAN CHEN** (bigbird.chenyan@huawei.com) received her B.Sc. and Ph.D. degrees in 2004 and 2009, respectively, from Zhejiang University. She has been a visiting researcher in HKUST during the year 2008 to 2009. She joined Huawei Technologies (Shanghai) in 2009 and has been the project leader of Green Radio research from 2010 to 2013. She is now technical leader of multiple access research and standardization. She has won the award for IEEE Advances in Communications in 2017.

**ALIREZA BAYESTEH** (Alireza.Bayesteh@huawei.com) received his Ph.D. in Electrical and Computer Engineering from University of Waterloo, Waterloo, Canada in 2008. Since 2011, he has been with Huawei Canada, Ottawa, where he is currently a staff engineer. His research interests include 5G wireless communications with focus on NoMA.

**YIQUN WU** (wuyiqun@huawei.com) received his Ph.D. degree in electronic engineering from Tsinghua University, China, in 2012. Since 2012, he has been with Huawei Technologies, Shanghai, China. His research interests include energy-efficient wireless networks, new waveforms, and multiple access schemes for 5G.

**BIN REN** (renbin@catt.cn) received his M.S. and Ph.D. degrees from Beijing University of Posts and Telecommunications, China, in 2009 and 2017, respectively. Since 2009, he was with the Key Laboratory of Wireless Mobile Communications, China Academy of Telecommunication Technology, Beijing, China. His research interests include pattern division multiple access, no-orthogonal multiple access and 5G wireless communications system. Till now he has published 10 academic papers, and applied more than 32 patents.

**SHAOLI KANG** (kangshaoli@catt.cn) received her Ph.D. in electrical engineering from Beijing Jiaotong University, China. She joined Datang Telecom Group in 2000 doing research on TD-SCDMA. She was a research fellow in University of Surrey, UK during 2005-2007. Since September 2007, she has been working in Datang, focusing on 4G and 5G technologies. She was in charge of China 863 5G project "R&D on 5G novel modulation and coding technologies".

**SHAOHUI SUN** (sunshaohui@catt.cn) received his Ph.D. from Xi'dian University, China, in 2003, and postdoctoral fellow with the China Academy of Telecommunication Technology, China, in 2006. Since 2011, he has been the Chief Technical Officer with Datang Wireless Mobile innovation Center of the China Academy of Telecommunication Technology. He is involved in the development and standardization of the 3GPP LTE and 5G. His research area of interest includes multiple antenna technology, heterogeneous wireless networks and NOMA.

**QI XIONG** (q1005.xiong@samsung.com) received his Ph.D. in Electrical & Electronic Engineering from the NANYANG Technological University, Singapore. He joined Communication Research Lab in Beijing Samsung Telecommunication R&D center as a 5G research engineer in 2015. His research interests include physical layer security, non-orthogonal multiple access, 5G communication PHY/MAC design etc. Currently, he is involved in standardization for 5G in 3GPP.

**CHEN QIAN** (chen.qian@samsung.com) received his Ph.D. in Electrical Engineering from the Tsinghua University, China. He joined Communication Research Lab in Beijing Samsung Telecommunication R&D center as a 5G research engineer in 2015. His research interests include MIMO system, waveform design, non-orthogonal multiple access, 5G communication PHY/MAC design etc. Currently, he is involved in standardization for 5G in 3GPP.

**BIN YU** (bin82.yu@samsung.com) received his M.S. in Electrical Engineering from the University of Southampton, United Kingdom. He joined Communication Research Lab in Beijing Samsung Telecommunication R&D center as a 5G research lab leader in 2013. His research interests include MIMO system, waveform design, non-orthogonal multiple access, 5G communication PHY/MAC design etc. Currently, he is involved in standardization for 5G in 3GPP.

**ZHIGUO DING** (z.ding@lancaster.ac.uk) received his Ph.D. degree from Imperial College London in 2005, and is currently a chair professor at Lancaster University, United Kingdom. His research interests include 5G communications, MIMO and relaying networks, and energy harvesting. He served as an Editor for several journals including IEEE Transactions on Communications, IEEE Communication Letters, IEEE Wireless Communication Letters, and Wireless Communications and Mobile Computing.

**SEN WANG** (wangsenyjy@chinamobile.com) received the Ph.D. degree in information and communication engineering from Beijing University of Posts and Telecommunications (BUPT), Beijing, China, in 2013. After graduation, he joined the Green Communication Research Center (GCRC), China Mobile Research Institute, as a Project Manager. His research interests include 5G air interface technologies, especially on MIMO, multiple access, radio resource allocation and performance evaluation for future cellular networks.

**SHUANGFENG HAN** (hanshuangfeng@chinamobile.com) graduated from Tsinghua University, Beijing, China, in 2006 and is now a senior project manager in green communication research center of china mobile research institute, leading 5G R&D. He is also vice chair of wireless technology work group of China's IMT-2020 (5G) promotion group. His research interests are mainly focused on 5G wireless communication systems, including massive MIMO, flexible duplex, NOMA, high speed train communication and wireless big data with AI.

**XIAOLIN HOU** (hou@docomolabs-beijing.com.cn) received his Ph.D. in communication and information system from Beijing University of Posts and Telecommunications, China. He joined DOCOMO Beijing Laboratories in 2005 and now he is the deputy director of wireless technology department. He has been actively contributing to 4G and 5G research, standardization and trials. His current research interests include massive MIMO, mmWave, flexible duplex, NOMA, URLLC and cellular V2X.

**HAO LIN** (hao.lin@orange.com) received his Ph.D. degree in communication and electronics from the Ecole National Supérieure des Télécommunications (ENST) Paris, France, in 2009. Since 2010, he has joined Orange Labs (Rennes) as a research engineer. His research interests include multi-carrier modulation and signal processing for communications. He has been leading several European research projects, including FP7 and 5GPPP. He is now involved in standardization for 5G representing ORANGE as a 3GPP RAN1 delegate.

**RAPHAEL VISOZ** (raphael.visoz@orange.com) received his Ph.D. degree in Digital Communications from the Ecole Nationale Supérieure des Télécommunications (ENST), Paris, France, in 2002. Since November 1997, he has been working for Orange in the field of 3/4/5G mobile radio systems. His research interests include network information theory, PHY/MAC cross layer optimization mechanisms, multi-antenna technology (MIMO systems), iterative decoding on graphs. He is now involved in standardization for 5G representing ORANGE as a 3GPP RAN1 delegate.

**RAZIEH RAZAVI** (razieh.razavi@vodafone.com) received her Ph.D. degree in mobile communications from University of Surrey, U.K., in 2012. She continued her research as a Research Fellow with the Institute for Communication Systems (ICS), Home of the 5G Innovation Centre, University of Surrey. Since November 2015, she has joined Vodafone Group's R&D team, focusing on 4G and 5G technologies. Her research interests include 5G wireless communications system, non-orthogonal multiple access, advanced multi-user detection and decoding techniques.